\documentclass[11pt,a4paper]{article}
\usepackage[left=2.3cm,top=2.3cm,right=2.3cm,nohead,nofoot]{geometry}

\begin{document}
%\pagenumbering{empty}
\title{Reply to comment on {\it Twisted protein aggregates and disease} by Liu and Gunton}
\author{M. S. Turner\\
 Department of Physics, \\
University of Warwick, Coventry CV4 7AL.\\
\tt{m.s.turner@warwick.ac.uk} }
\date{\today}
\maketitle
In a recent post on arXiv \cite{liu} Liu \& Gunton comment on the role of a bulk chemical potential~$\epsilon$ in determining the radius $R$ of a self-assembled bundle of chiral protofilaments of indeterminate length $L\gg R$. In essence their concern was that our work \cite{theorypaper} erroneously omitted the effect on $R$ of a bulk chemical potential representing exchange between the solvated pool of monomers and the polymerised bundle (polymer).

We believe that Liu \& Gunton's criticism is fundamentally flawed because they work in the ensemble of fixed $\epsilon$ and, implicitly, {\bf fixed polymer length}. In this ensemble $\epsilon$ is indeed conjugate to $R^2$ and so can control the polymer radius. However, the physically relevant ensemble is one in which the  {\bf polymer length is free to adjust}. Here $\epsilon$ is conjugate to $L$ and the aggregate can grow by elongating at a fixed equilibrium radius $R$, corresponding to the minimum polymer free energy density, here denoted $\epsilon_{{\rm pol}}$. When polymers form the monomeric concentration falls and the chemical potential adjusts until, finally,  $\epsilon=\epsilon_{{\rm pol}}$ and only polymers with this equilibrium radius are (marginally) stable. Thus, whenever polymers form, {\bf the final value of $\epsilon$ is not a free parameter.} There is a simple physical argument for why the radius of the aggregate must be insensitive to its length, and hence to the total mass of aggregated polymer: In the absence of long ranged interactions there is no physical coupling between $R$ and~$L$.

Since our original letter other authors have undertaken similar studies in which a bulk chemical potential was also included \cite{nottotallystupidquestions}. These authors take the limit in which the polymer length $L\to\infty$, eliminating what is otherwise an {apparent} dependence of $R$ on $\epsilon$, e.g. in Eq 7 in \cite{nottotallystupidquestions}. 
%These authors claim that the {\it variation} of the radius $\delta R\sim L^{-1/2}$ which we find hard to reconcile with the above physical argument.

In the interests of completeness we would add that the chemical potential {\it can} affect the aggregate radius but only if the concentration is either near the CMC, where the aggregates are so small as to no longer be essentially linear, or so high that strong polymer-polymer interactions are important. Neither regime is the focus of \cite{liu}, \cite{theorypaper} or \cite{nottotallystupidquestions}. 

For those with an interest in the total polymer length, equivalent to the polymer concentration, we add that this could be calculated from $\epsilon_{{\rm pol}}$ given in \cite{theorypaper} (or \cite{nottotallystupidquestions}), combined with an appropriate relationship between $\epsilon$ and the concentration of monomers in solution $\phi$. The final monomer concentration is given by $\epsilon(\phi)=\epsilon_{{\rm pol}}$ and the corresponding increase in the amount of polymer follows by conservation of material.
\bibliographystyle{prsty.bst} %unsrt
\bibliography{guntonreply}

\end{document}